\documentclass[12pt,aps,prl, onecolumn, superscriptaddress, notitlepage]{revtex4-2}
\usepackage{amssymb,amsmath}
\usepackage{natbib}
\usepackage{graphicx}
\usepackage{amsthm}
\usepackage{cancel}
\usepackage{color}
\usepackage{bbold}
\usepackage{wrapfig}
\usepackage{booktabs}
\usepackage{float}
\usepackage{enumitem}
\usepackage{hyperref}
\usepackage{tikz}
\usepackage{comment}
\usepackage{longtable}
\usepackage{multirow}
\usepackage{hyperref}

\usepackage[normalem]{ulem}

\begin{document}

\title{Disentangling Scaling Arguments\\ to Empower Complex Systems Analysis}

\author{Marc Timme}
\affiliation{Chair for Network Dynamics,\\
Institute for Theoretical Physics, Center for Advancing Electronics Dresden (cfaed),  Center of Excellence Physics of Life, Technical University of Dresden, Germany}
\author{Malte Schr\"oder}
\affiliation{Chair for Network Dynamics,\\
Institute for Theoretical Physics, Center for Advancing Electronics Dresden (cfaed),  Center of Excellence Physics of Life, Technical University of Dresden, Germany}

\date{\today}

\maketitle

\textbf{Scaling arguments provide valuable analysis tools across physics and complex systems yet are often employed as one generic method, without explicit reference to the various mathematical concepts underlying them. A careful understanding of these concepts empowers us to unlock their full potential.}

The notion of scaling plays a central role across physics as well as complex systems analysis. 
It offers powerful tools for answering fundamental questions about a system without considering all its details. 
However, scaling arguments are often used without careful consideration of their various distinct mathematical definitions. 
Yet, their full potential can be exploited only by respecting these different concepts.
We may be able to classify system structure and dynamics, to understand mechanisms underlying (collective) phenomena complex systems display, and, sometimes, to quantitatively predict system behavior. \\

In Statistical Physics, scaling exponents naturally appear in the analysis of continuous phase transitions.
Here, an observable of the system -- an order parameter -- quantifies properties of the system that change between two qualitatively different macroscopic states. 
At a critical value of an external control parameter such as temperature, the order parameter continuously grows from zero, indicating the emergence of a partially ordered state. 

Near the phase transition point, the order parameter typically grows as a power law with the distance from that critical
point. 
For instance, in ferromagnetic systems, as characterized by the Ising model \cite{ising1925_beitrag, mccoy2014_two}, the macroscopic magnetization is zero above some critical temperature. The magnetic moments (spins) are not aligned and thus do not induce macroscopic magnetization. 
Below the critical temperature, the system's symmetry is broken as spins align with each other macroscopically in one preferred spatial direction. 
The overall magnetization of the system changes continuously from zero to non-zero values at the critical temperature. 
In the mean field solution of the Ising model, the average magnetization in the ordered phase grows with the square root of the distance to the critical temperature. 
In this sense, power law scaling constitutes an essential indicator of continuous phase transitions, with the value of the scaling exponent distinguishing between different subtypes of transitions. 

Intriguing theoretical and empirical insights starting in the 20th century demonstrated that these critical exponents characterizing continuous phase transitions often are universal. 
Their value is largely independent of the material properties but instead characterizes the similarity of the underlying mechanisms and the effective dimensionalty of the underlying interactions across different systems. For our example of ferromagnetic systems,
these critical exponents essentially depend on the dimension of the system and the degrees of freedom of the magnetic moments, yet not on the atoms or molecules carrying these moments or details of their interactions. 
Moreover, the same mean field exponent 1/2 is shared with a host of entirely different systems and phenomena \cite{goldenfeld2018_lectures}. As a result of this universality, how order emerges and thus how order parameters scale near critical points enables us to classify these systems by the scaling exponents they exhibit. 

\begin{figure}
    \centering
    \includegraphics{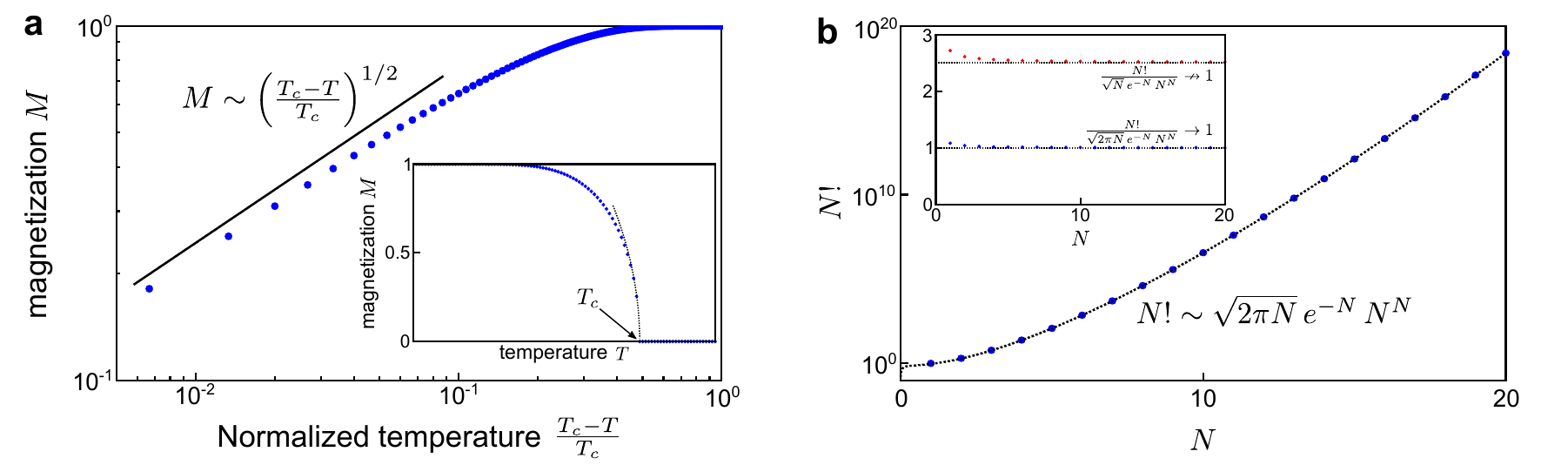}
    \caption{\textbf{Two distinct types of scaling analysis.} Universal power law scaling in continuous phase transitions (panel a) and asymptotic expansions (panel b) constitute two powerful tools of scaling analysis for physics and complex systems. We emphasize that the two panels illustrate two distinct notions of scaling, each with a different meaning of the symbol  ``$\sim$'' according to  Eqns.~\eqref{eq:PowerLawExponent} and \eqref{eq:AsymptoticLimit}, respectively.}
    \label{fig:scaling}
\end{figure}

Although (universal) scaling exponents are key to characterize and classify different systems, there is more to power law scaling than the exponent. 
Even if the exponent of a power law is known, its knowledge is insufficient to quantitatively predict the value of an order parameter (or any other quantity exhibiting the power law scaling). 
Unfortunately, unclear and inconsistent mathematical notation employed across physics and complex systems analysis further mystifies the issue. 
To characterize the square root scaling of the magnetization $M$ with the temperature difference $T_{\text{c}}-T$ to a critical temperature $T_{\text{c}}$ in ferromagnets, we often write $M\sim(T_{\text{c}}-T)^{\beta}$. 
Mathematically, this notation is short hand for 
\begin{equation}
\lim_{\varepsilon\rightarrow0+}\frac{\ln M(\varepsilon)}{\ln\varepsilon} = \beta \label{eq:PowerLawExponent}
\end{equation}
where $\varepsilon= \left(T_\text{c}-T\right)/T_\text{c}$ quantifies the normalized distance to the critical temperature (Fig~\ref{fig:scaling}a). 
On first glance, it may appear as if we could theoretically predict $M$, at least approximately, for small temperature differences $(T_{\text{c}}-T)$ given the exact definition (\ref{eq:PowerLawExponent}) if we know the exponent $\beta$ (along with the critical temperature $T_{\text{c}}$). 
However, we are missing an overall constant. 
If $M=A\times\left[\left(T_\text{c}-T\right)/T_\text{c}\right]^{\beta}$, also the constant $A$ determines the actual value of $M$ for $T<T_{\text{c}}$. 
Without it, no prediction is possible. 
In contrast to the scaling exponents, the value of $A$ is not universal and typically depends on detailed material properties. For instance, the prefactor $A$ may depend not only on the type of spin and the structure of the interactions but also on the specific type of atoms carrying these spins. 
The same holds for all scaling laws across systems, settings, and subjects \cite{grimmet_percolation, west2017scale, d2015anomalous, newman2005power, clauset2009power}: 
The scaling exponent classifies the dynamics and phenomena and enables qualitative estimates of how observables change if control parameters vary, but is insufficient to quantitatively predict absolute magnitudes of observables.\\

More general types of scaling may be revealed by asymptotic analysis where systematic expansions of an observable function $f$ near a point $x_{0}$ do yield quantitative predictions for observables. 
Already undergraduate courses in Mathematics teach us the usefulness of Taylor series expansions of a function to estimate its value close to some point of interest. 
For instance, the third-order Taylor polynomial of the sine function, $\sin(x)=x-x^{3}/6+\ldots$ near $x_{0}=0$, provides reasonable estimates of it as long as $|x|$ is sufficiently small. 
Taylor's theorem \cite{taylor1715_methodus, altland2019mathematics} moreover provides explicit bounds on the error we make if we approximate a function by a finite-order Taylor polynomial in a given interval. 
More advanced approaches yield asymptotic expansions $g(x)$ that become asymptotically equal to a give function $f(x)$ as $x\rightarrow x_{0}$. 
In asymptotic analysis, a subfield of Mathematics, the notation $f(x)\sim g(x)$ as $x\rightarrow x_{0}$ denotes that 
\begin{equation}
\lim_{x\rightarrow x_{0}}\frac{f(x)}{g(x)}=1,\label{eq:AsymptoticLimit}
\end{equation}
signifying that the function $g$ serves as an increasingly accurate approximation of $f$ the closer $x$ is to $x_{0}\,$.  
As opposed to Taylor series, asymptotic series often do not converge and may exhibit intricate, non-standard prefactors \cite{bender2013_advanced, miller2006_applied}. Such asymptotic series are common in physics and complex systems analysis. 
For instance, the Wentzel-Kramers-Brillouin (WKB) approximation \cite{bender2013_advanced} relies on asymptotic analysis to provide semi-classical estimates of quantum wave functions resulting from a given potential. 
The asymptotic nature of such expansions is often not explicated. 
Rather, even many text books and lecture notes often consider it some (commonly not further defined) variant of Taylor series and neither mention convergence aspects nor non-standard prefactors. 

So asymptotic expansions, as Taylor expansions, help to accurately predict the value of a function close to a point $x_{0}$ of interest by providing a partial sum of a (convergent or divergent) series that becomes closer and closer to the actual function as $x\rightarrow x_{0}$. 
Although formally only slightly different, asymptotic expansions are vastly more powerful in making quantitative predictions than knowing powers of a power law, even if the power law does become exact near $x_{0}$ and the exponent is known analytically. 

To clarify the issue, let us consider a simple, purely mathematical example. 
The factorial $N!$ of a large integer $N$ is difficult to handle analytically. 
A commonly known approximation is provided by Stirling's formula, $N!\sim\sqrt{2\pi N}e^{-N}N^{N}$ as $N\rightarrow\infty$ which constitutes the lowest-order term of
an asymptotic expansion. 
The exact mathematical meaning of this scaling relation is that the quotient $N!/(\sqrt{2\pi N}e^{-N}N^{N})$ approaches $1$ as $N\rightarrow\infty$ in the sense of eqn.~(\ref{eq:AsymptoticLimit}) (Fig~\ref{fig:scaling}b). 
Thus, viewed as a statement about asymptotic scaling, an alternative formula $N!\sim\sqrt{N}e^{-N}N^{N}$ would be incorrect. 
The ``approximation'' would be too small by a factor of $\sqrt{2\pi}=2.50\ldots$ and thus yields no accurate quantitative prediction. 
Similar challenges occur for simple power laws. 
Let us consider, as in the example above, that $M=A\,\varepsilon^{\beta}$ is some order parameter, $\varepsilon$ the distance to a transition point, and $A$ some prefactor. 
Finding the exponent $\beta$ via (\ref{eq:PowerLawExponent}) is thus insufficient to quantitatively predict $M$ and constitutes only the first step of analysis. 
To predict the value of $M$ at any point where $\varepsilon > 0$ we need a second step
to find 
\begin{equation}
\lim_{\varepsilon\rightarrow0^+}\frac{M(\varepsilon)}{\varepsilon^{\beta}}=A.
\end{equation}
In the language of asymptotic analysis, we then discovered that $M(\varepsilon) \sim A\,\varepsilon^{\beta}$ as $\varepsilon \rightarrow 0^+$ with a well-defined prefactor $A$. Only with this factor, predicting $M$ becomes possible.\\

At the turn of the century, Stephen Hawking stated that the 21st century is the century of complexity. 
For phenomena substantially more complex than ferromagnetic ordering, employing arguments of power law scaling or asymptotic scaling (or both) typically is much more involved, not least because the consequences of the scaling of some observable are often indirect. They need to be evaluated in convoluted, nonlinear ways to yield conclusions about the original quantity of interest. 
For instance, work by Hens et al. \cite{hens2019spatiotemporal} unraveled universal features of signal propagation patterns in networks of dynamical elements that result from permanently fixing the state of a single unit to a new value. 
They find qualitatively different types of propagation patterns \cite{timme2019propagation}: some whose spreading properties are largely controlled by the graph-theoretical distance from the perturbed unit and others where spreading strongly depends on how many other units the intermediate units interact with. 
In a recent Comment \cite{peng2020_comment}, Peng et al. argue that under certain conditions, a prefactor in a central expression is missing and that this prefactor is important for predicting propagation times. 
At the same time, the Reply \cite{hens2020_reply} by Hens et. al argues that common scaling analyses in Statistical Physics are made to neglect prefactors and focus on scaling exponents because these reveal whether certain collective features are universal. 

Intriguingly, both the authors of the original article (and the Reply) and those of the Comment may be simultaneously correct. 
The conflict in argument likely originates from a different interpretation of the notion of scaling as well as a different notational meaning the respective author groups assign to the mathematical symbol ``$\sim$''. 
Finally, their interpretation depends on which aspects of the analysis each group considers important. 
As detailed above, scaling arguments invoked for extracting scaling exponents are useful for characterizing system dynamics and their universality across different settings. 
In contrast, asymptotic scaling analysis is capable of much stronger statements and, in addition to revealing scaling exponents, helps approximately predicting the absolute
magnitude of a quantity of interest. 
Roughly speaking, estimates of scaling exponents help to predict key qualitative features of observable behavior whereas asymptotic analysis helps to make quantitative predictions of observables. 

Fully embracing the different aspects of scaling
thus helps analyzing and predicting phenomena across physics and complex systems.
The tools of scaling analysis become particularly powerful once we exactly distinguish the different notions of scaling and their distinct implications.

\end{document}